# ADEPT Drag Modulation Aerocapture: Applications for Future Titan Exploration

Athul Pradeepkumar Girija [1,**]

[1]School of Aeronautics and Astronautics, Purdue University, West Lafayette, IN 47907, USA

## ABSTRACT

The Cassini-Huygens mission has transformed our understanding of Titan from a hazy veiled moon to a place surprisingly like the Earth, with terrestrial physical processes such as wind, rainfall, and erosion shaping the landscape albeit with entirely different chemistry and temperatures. Dragonfly, a single element mission which fits within the New Frontiers cost cap will arrive at Titan in 2034, and perform in-situ investigations of the organic materials on the surface. However, its detailed investigations will be limited to region within its short flight range. The big gaps in our understanding of Titan's global topography, climate, and upper atmospheric chemistry which can only investigated from an orbiter around Titan will remain to be addressed by a future orbiter mission. Due to the challenges of attaining orbit, past Titan orbiter concepts have been beyond the New Frontiers cost cap. The present study explores the use of drag modulation aerocapture for a Titan Orbiter which fits within New Frontiers. The study shows how a Dragonfly-like lander, and a Titan orbiter which each inividually fit within the New Frontiers cost cap, when combined together can provide the science data return equivalent to a Flagship-class mission.

---

[**] To whom correspondence should be addressed, E-mail: athuplg007@gmail.com

# I. INTRODUCTION

Saturn's largest moon, Titan, is the only moon in the Solar System that has a significant atmosphere, and has been a prime target for scientific investigations since the Pioneer 11 flyby in 1979 [1]. In 1980, the Voyager 1 spacecraft performed the first close flyby of Titan, and radio occultation experiments provided the first measurements of the vertical temperature and pressure profiles of the atmosphere down to the surface [2]. However, Titan's surface was essentially hidden from view by the thick atmospheric haze, and would remain a mystery until the arrival of the Cassini-Huygens spacecraft in 2004. The Huygens probe entered the Titan atmosphere, and accomplished a successful landing near the equator, making the first in-situ measurements of the atmosphere and of the surface. The probe also returned the first images from Titan's surface, revealing a landscape strewn with rounded pebbles made of water-ice and indications of past fluvial activity [3]. Over the next several years, the Cassini spacecraft performed several close flybys of Titan. Using its radar to see through the thick haze, Cassini revealed the presence of hydrocarbon lakes on Titan's surface, making it the only planetary body known to harbor surface liquids other than the Earth [4]. Cassini also provided evidence that Titan's surface is highly varied with sand dunes near the equator made of organic dust of unknown composition, and a large number of lakes and seas near the poles made of methane and ethane [5]. Cassini also provided evidence that Titan has an active hydrological cycle like the Earth, with methane instead of water, resulting in the formation of clouds, seasonal storms, and rainfall leading to fluvial patterns and erosion [6]. Cassini revealed that Titan was one of the most scientifically interesting places in the Solar System, with its thick atmosphere which may resemble that of the early Earth, a surface coated with organic material falling from the atmosphere which over billions of years may have had the chance to interact with liquid water from cryo-volcanism or impact craters, and likely a subsurface liquid water ocean in its deep interior [7]. Each of these unique realms of Titan and their interactions make it a very compelling scientific target for understanding how simple organic compounds can evolve into more complex systems which may hold clues of the origin of life on the early Earth [8]. Even as the Cassini mission was just beginning its science mission, a number of follow on mission concepts were being formulated to further investigate the early discoveries at Titan. One of the first concepts was the Titan Prebiotic Explorer (TiPEx) in 2006, consisting of a two-element mission: an orbiter for global observations and data relay, and a montgolfier hot air balloon for in-situ measurements of the atmosphere and the surface [9]. The theme of using an orbiter and in-situ elements both for synergy of global and local investigations, and using the orbiter to relay the data from the in-situ element such as a lander or balloon would be recurring one seen in several future Titan mission concept studies.

In 2007, the Titan Explorer Flagship mission study proposed a three element mission consisting of an orbiter, a montgolfiere balloon, and a lander with an estimated cost of $3.5B (FY 2007) [10]. In 2009, the joint NASA-ESA Titan Saturn System Mission study proposed a similar concept as the Titan Explorer but without the use of aerocapture for the orbiter, instead using a Solar Electric Propulsion (SEP) stage during the interplanetary cruise to reduce the encounter velocity for chemical propulsive capture [11]. The estimated mission cost was about $4B (FY 2007). By the late 2000s, the 2013-2022 Planetary Science Decadal survey recommended the Europa orbiter for the Flagship class mission for the next decade and the Titan Flagship mission plans were essentially shelved. However, in 2010, two smaller Discovery-class (< $500M) mission concepts for Titan were proposed. The Titan Mare Explorer (TiME) proposed a floating lander on Titan's northern seas [12]. However due to the timing of the mission, Titan's northern seas would be in permanent darkness during the winter during its arrival and render direct-to-Earth (DTE) communication impossible. A relay orbiter was out of reach within the Discovery-budget and proposal was eventually not selected. The AVIATR concept proposed an airplane which would fly non-stop around Titan, but the cost estimate proved to be infeasible with the Discovery-class budget [13]. After a period of relative inactivity, in the late 2010s, a major breakthrough occurred when NASA selected the Dragonfly mission for flight under the New Frontiers program under $900M. The Dragonfly concept is a single-element mission: a quad-copter which is essentially a relocatable lander [14]. By combining the elements of previously proposed montgolfier balloons and lander into a single element, and using DTE communication (thus removing the need for an orbiter), Dragonfly was able to significantly reduce the mission cost compared to previous Flagship proposals and fit within the New Frontiers cost cap. The mission is expected to launch in 2027, and will arrive at Titan in 2034. Dragonfly will land in the equatorial dune fields near the Selk impact structure which also has evidence of cryo-volcanic activity in the region nearby. The location is of great interest for in-situ exploration as the surface organics may have come in contact with liquid water, thus providing a unique opportunity to sample these organic materials [15]. While Dragonfly is without question one of most exciting scientific missions planned to date, its detailed investigations will be limited to region within its short flight range. The big gaps in our understanding of Titan's global topography, climate, and upper atmospheric chemistry which can only investigated from an orbiter around Titan will remain to be addressed by a future mission [16]. In addition, the cost-saving DTE approach comes with a penalty, as the data rates are severely limited. The present study explores the possibility of using drag modulation aerocapture with a deployable entry system, a technique which uses atmospheric drag to insert an orbiter around Titan with very little propellant. Titan's thick extended atmosphere and low-gravity make it the most attractive destination in the Solar System for performing aerocapture.

## II. DRAG MODULATION AEROCAPTURE

Orbit insertion of spacecraft at planetary destinations is a maneuver which requires substantial velocity change ($\Delta V$). Traditionally, this $\Delta V$ is achieved using a chemical propulsion rocket engine which decelerates the spacecraft and allows it to be captured into orbit around the planet. The Cassini spacecraft for example, performed a 633 m/s burn for its orbit insertion into a highly elliptic orbit around Saturn. For a future Titan mission, it would be desirable to have an orbiter around Titan and not Saturn to accomplish its high resolution global mapping goals. However, inserting an orbiter around Titan, particularly to a low-circular orbit requires a $\Delta V$ of about 4+ km/s, which is very challenging for chemical propulsion systems due to the large amount of propellant needed. A promising alternative solution is to use Titan's atmosphere to decelerate the spacecraft with aerocapture and has been studied in detail for over two decades. The most prominent is the NASA Titan Aerocapture Systems Analysis in 2003, which performed a detailed study of using a low-lift-to-drag (L/D) ratio blunt-body aeroshell to insert an orbiter into a 1700 km circular orbit around Titan [17]. The results from the study were leveraged in the TiPeX and Titan Explorer Flagship proposals. While the low-L/D blunt-body aeroshells have extensive flight heritage from Mars missions, and no new technology developments were needed, Titan missions using these aeroshells would still be expensive. For reference, a de-scoped version of the Titan Explorer which removed all the in-situ elements and retained only the orbiter was still estimated to cost $2B (FY 2007), nearly twice the New Frontiers cost cap. One of the reasons for the high cost is that the lifting aeroshell design requires high rate reaction control thrusters to roll the vehicle during the flight to control its trajectory, as well as use ballast masses which are jettisoned before entry to offset the center-of-gravity from symmetry axis giving it the required L/D at a constant angle-of-attack. In 2014, Putnam and Braun proposed the use of drag modulation flight control for aerocapture to overcome these challenges associated with lifting blunt-body aeroshells, and thus potentially reduce the cost of aerocapture missions with a simpler control technique [18]. The vehicle has no lifting capability (L/D = 0) and the only control variable is the drag area. There are several variants of the drag modulation technique, but the simplest is the single-event jettison concept. In this method, the vehicle enters the atmosphere with a large drag area and then jettisons the drag skirt at the appropriate time when enough speed reduction has been achieved using atmospheric drag [19]. The vehicle then flies the rest of the atmospheric trajectory with the small drag area and exits the atmosphere, after which it performs a small periapsis raise maneuver at the apoapsis and achieves its initial orbit. The concept of operations for drag modulation aerocapture is shown in Figure 1. Drag modulation aerocapture has been extensively studied in recent years for its applications to inserting small satellites into orbit around Mars and Venus, which is very challenging for small satellite propulsion systems [20].

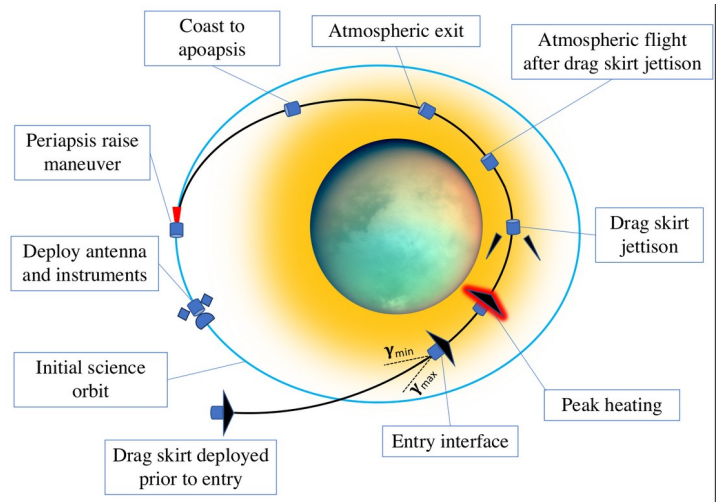

Figure 1. Concept of operations for drag modulation aerocapture.

### III. ADEPT DRAG MODULATION AEROCAPTURE SYSTEM

The Adaptable, Deployable, Entry and Placement Technology (ADEPT) is a deployable entry system developed by NASA for a wide range of future planetary missions and destinations [21]. It is essentially folded like an umbrella during launch, and deploys to its full diameter before entry. This architecture is scalable from small sub 1-m entry systems for small satellites to large 12-m diameter systems for large missions, and provides several advantages over conventional blunt-body aeroshells. First, its small diameter when stowed implies it can be easily accommodated within the available diameter of a launch vehicle fairing. Second, its large large diameter when deployed gives it a low ballistic coefficient which enables the deceleration to occur higher up in the atmosphere and thus considerably lowers the peak heating compared to a rigid aeroshell. This avoids the need for ablative heat shield materials, and the carbon-cloth fabric which makes up the drag skirt doubles as both the structural element and the thermal protection system. Third, jettison of the drag skirt during flight provides the necessary change in the vehicle drag area and thus enables ADEPT to be used as a drag modulation flight control system without the need for reaction control thrusters. Figure 2 shows the ADEPT Sounding Rocket (SR-1) flight test article in its showed and deployed configurations [22].

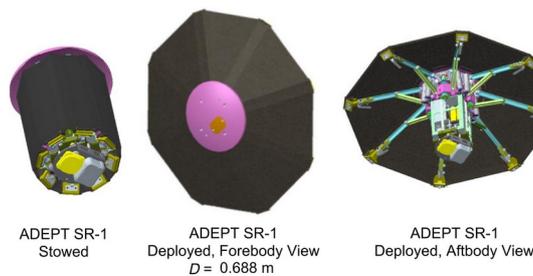

Figure 2. Schematic of the ADEPT SR-1 flight test article in its stowed and deployed configurations [22].

Recent studies have investigated the use of ADEPT drag modulation aerocapture for insertion of small satellites, including constellations of small satellites into orbits around Mars and Venus [23]. To date, most drag modulation aerocapture studies have focused on Mars and Venus. However, drag modulation aerocapture is not just limited to small missions at Mars and Venus. It is also applicable for large missions to the outer Solar System, though outer planet aerocapture studies are mostly limited to lift modulation architectures [24, 25, 26]. There are very few studies which address the performance of drag modulation aerocapture at the outer planets, and their implications for future missions. In 2021, a study by Strauss et al. investigated the use of a large ADEPT drag modulation system for a Neptune orbiter [27]. Figure 3 shows a schematic of a 12-m ADEPT concept from the study. To date, there have been no studies investigating the use of ADEPT drag modulation aerocapture for large orbiters at Titan. In the present study, the performance of the ADEPT system for aerocapture at Titan for a large orbiter is investigated, along with its scientific and engineering implications for future Titan New Frontiers and Flagship mission concepts.

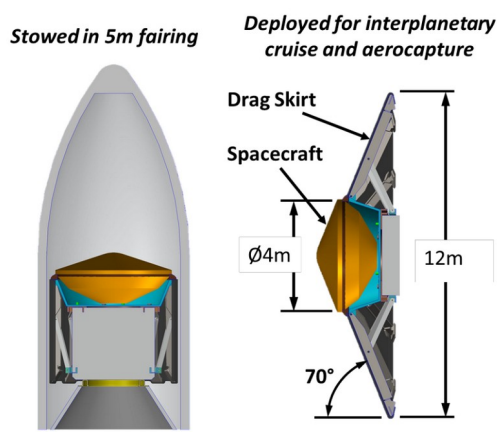

Figure 3.  Schematic of a 12-m diameter ADEPT concept for large outer planet missions [27].

## IV. AEROCAPTURE PERFORMANCE

Due to the limited scope of the study, a broad trajectory search to identify promising potential interplanetary trajectories to Saturn is not performed. Instead, the baseline interplanetary trajectory used by the Dragonfly mission is used as a reference [28]. The trajectory will launch in June 2027, and use a VEEEGA (V=Venus, E=Earth) gravity assist trajectory to arrive at Saturn in December 2034. The hyperbolic excess speed at arrival (with respect to Titan) is 6.98 km/s. While Dragonfly will target a direct entry to deliver the lander to Selk crater near the equator, the proposed orbiter mission will target an entry for aerocapture to achieve a near-polar orbit around Titan.

The entry speed at atmospheric interface (1000 km) is 7.34 km/s. The aerocapture vehicle design is the same 70-degree sphere cone as that used by Strauss et al. and shown in Figure 3. The vehicle entry mass is 5700 kg (with the 12-m ADEPT drag skirt), and the spacecraft mass after drag skirt jettison is 2600 kg. The ballistic coefficient of the entry configuration is 30 kg/m$^2$, and the ballistic coefficient ratio before and after drag skirt separation is 4.14. The vehicle nose radius is 1.0 m. The target orbit is 1700 km circular which is high enough above the atmosphere to prevent orbital decay, and yet close for radar measurements. The orbital inclination is 85 degrees for good coverage of the surface, and the off-polar inclination allows the orbit plane to be rotated by Saturn's gravity [29].

There are two key parameters which are used to quantify aerocapture performance for preliminary mission design: 1) the width of the aerocapture entry corridor, known as the Theoretical Corridor Width (TCW); and 2) the aero-thermal environment encountered by the vehicle, quantified by the peak stagnation-point heat rate, and the total heat load [30]. The TCW must be sufficiently large to accommodate errors in the entry-flight path angle (EFPA) $\gamma$ from navigation uncertainties, as well as atmospheric, and aerodynamic uncertainties. A detailed discussion of the calculation of the required TCW based on the various uncertainties is beyond the scope of the paper. A general rule of thumb for outer planet missions is to use a required TCW of at least 1 deg., to accommodate a ±0.3 deg. 3σ EFPA navigation uncertainty and a 0.4 deg. margin for atmospheric and aerodynamic uncertainties [31]. A smaller TCW can be used if the delivery errors, for example are smaller. The carbon-cloth thermal protection system (TPS) used in ADEPT has been tested to about 200 W/cm$^2$ heat rate and can accommodate total heat loads as high as 40 kJ/cm$^2$.

Using the entry speed and the vehicle design parameters stated above, the open-source Aerocapture Mission Analysis Tool (AMAT) is used to compute the shallow and steep limits of the aerocapture entry corridor [32]. The results are shown in Table 1. Titan's thick and extended atmosphere results in a steep entry corridor at [-36.6, -34.4] deg. compared to other destinations such as Mars or Venus where the corridor is much shallower. The width of the corridor is 1.89 deg., which is quite high considering the modest vehicle ballistic coefficient ratio of 4.14. For reference, the same vehicle at Neptune provides only 0.45 deg. of corridor. In fact, it has been shown that for a given drag modulation vehicle design and keeping other parameters such as the target orbit comparable, Titan provides the largest aerocapture entry corridor of any Solar System destination [33]. The corridor width is well above the 1 deg. requirement and provides adequate margin against the expected navigation and atmospheric uncertainties.

Table 1. Drag Modulation Aerocapture Entry Corridor at Titan

| Aerocapture corridor | Value, deg. |
|---|---|
| Shallow limit | -34.42 |
| Steep limit | -36.31 |
| TCW | 1.89 |

Figure 4 shows a nominal drag modulation aerocapture trajectory with a selected EFPA = -35 deg. near the middle of the entry corridor. The plots show the evolution of altitude, speed, deceleration, and stagnation-point peak heat rate as a function of time with t = 0 indicating the time at entry. It is worth pointing out that once again due to Titan's thick extended atmosphere, the duration of the manuever from atmospheric entry to exit is considerably long at 45 minutes. For other planets, the duration of the aerocapture maneuver is typically less than 10 minutes. The speed drops from 7.34 km/s at entry to 1.58 km/s at exit, resulting in an effective ΔV of 5.76 km/s. The peak deceleration is 3.4g, and the discontinuity in the deceleration at t = 6 min indicates the drag skirt jettison event. The peak stagnation point heat rate is 29 W/cm$^2$. The analysis only includes convective heating from the Sutton-Graves correlation. However, it is known that radiative heating is signficant at Titan and can be as much as the convective heating rate. Hence, an upper limit of 60 W/cm$^2$ is estimated for the stagnation point peak heat rate which is still well within the tested 200 W/cm$^2$ limit of the carbon cloth TPS. The total heat load (integral of the stagnation-point heat rate, multiplied by a factor of 2 to account for radiative heating) is about 10 kJ/cm$^2$, again well within the capability of the ADEPT entry system TPS. The nominal periapsis raise ΔV is 158 m/s, and the nominal spacecraft mass in orbit after aerocapture is 2500 kg. For comparison, if a chemical propulsion system with Isp = 320s was used to perform the orbit insertion with same arrival mass of 5700 kg, the mass in orbit would only be around 900 kg, not accounting for finite burn losses. Hence, drag modulation aerocapture is able to deliver 180% more mass to Titan orbit compared to propulsive insertion. The substantial reduction in propulsive ΔV required can translate into significant cost savings.

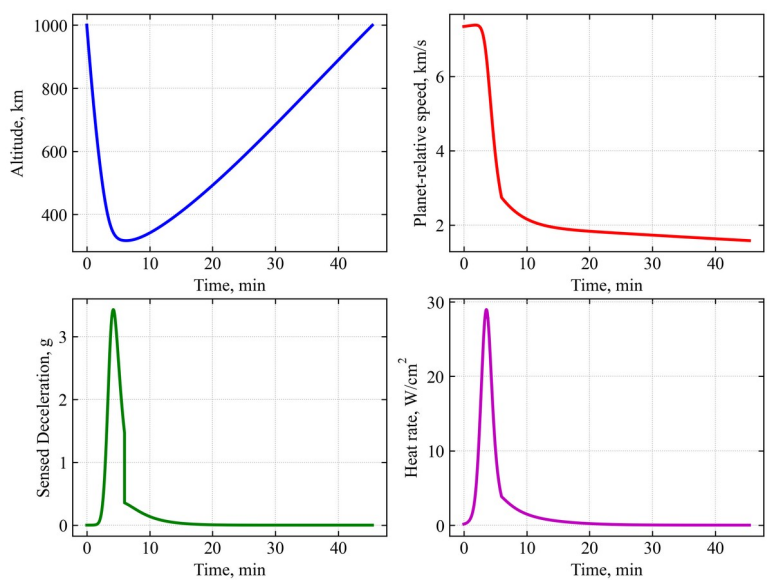

Figure 4. Evolution of the altitude, speed, deceleration, and heat rate for the aerocapture trajectory at Titan.

Figure 5 shows the 1700 km circular, 85 deg. inclination orbit at Titan after aerocapture. The transparent outer shell indicates the approximate extent of Titan's sensible atmosphere (500 km) above the solid surface which is indicated by the inner opaque shell. This schematic illustrates how thick Titan's atmosphere is in comparison to the surface of the planet (radius = 2575 km). The white ring indicates Titan's equatorial plane.

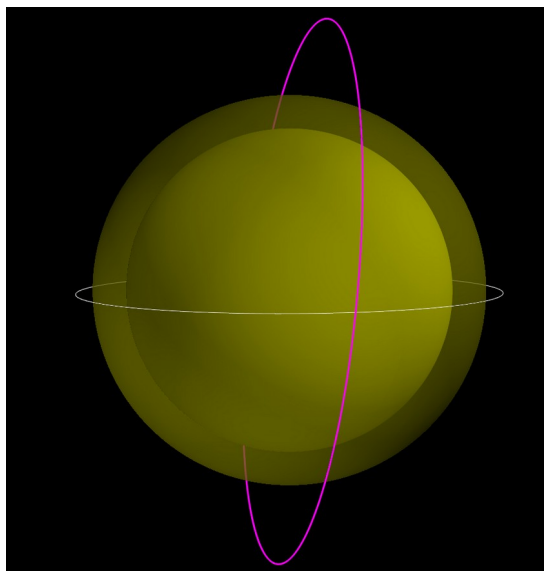

Figure 5. The 1700 km circular, 85 deg inclination orbit at Titan after aerocapture.

### V. ORBITER AS DATA RELAY FOR IN-SITU ELEMENTS

In addition to its primary goal of mapping and other global measurements of Titan, an orbiter around Titan can serve as high throughput data relay for in-situ elements such as a lander. There are two main reasons for this: 1) Lander antenna must fit within aeroshells and needs to be stowed during atmospheric flight. For example, High Gain Antenna (HGA) on Dragonfly has an approximate diameter of 1 meter. Orbiting spacecraft on the other hand typically carry much larger antennae. Cassini for example, carried a 4 meter HGA. Higher antenna diameter provides higher transmission gain and can thus enable higher data rate transmission. 2) As Titan rotates around its axis once every 16 days, a lander in the low latitudes will go behind the Earth for about 8 days during which no data transmission is possible. Depending on the Titan season, a lander in the high latitudes may be permanently visible from Earth or permanently not in view as seen in Figure 6. At the time of Dragonfly's arrival in 2034, its landing site at 3 deg N has Earth visibility for 8 days, and then Earth sets below the horizon. For a landing site at 80 deg N, the Earth is never visible as the north pole is in winter. For a landing site at 80 deg. S, the Earth is always above the horizon as the south pole is in permanent daylight. An orbiting spacecraft can overcome this problem for DTE and maintain communication with the in-situ elements irrespective of their location on the surface or the Titan season.

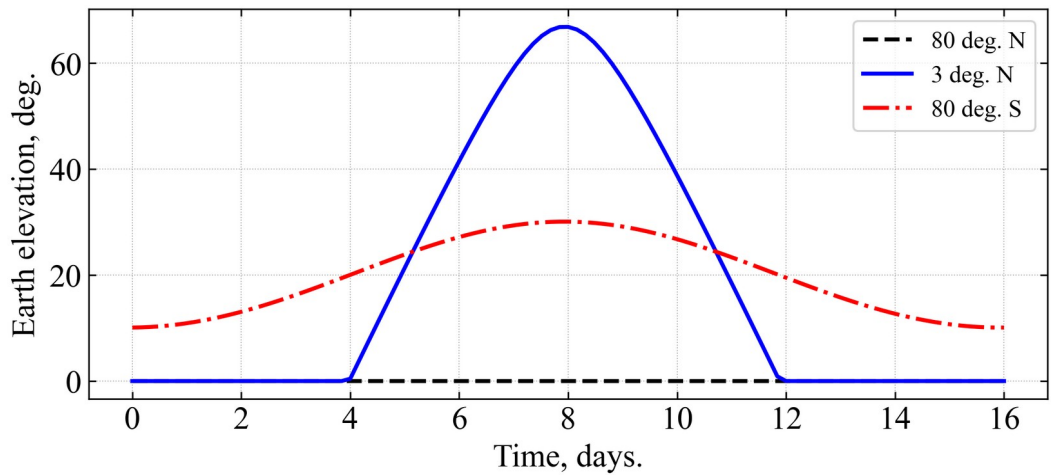

Figure 6. Elevation of Earth in the Titan sky for various latitudes during Dragonfly's arrival in 2034.

In addition to the above factors, in-situ elements cannot use the Ka-Band for DTE communications as using Ka-Band is susceptible to attenuation by the atmosphere at very low elevations. Orbiter spacecraft on the other hand, can use Ka-Band for higher rate transmission to the Deep Space Network (DSN). Hence if an orbital asset is available, the in-situ element can use a small antenna and X-Band to uplink its data to the orbiter, which would then use its large Ka-Band HGA to relay the data to DSN. This study performs a basic link budget and data volume analysis to quantify the science data return improvement from an in-situ element with an orbiting relay asset.

Table 2 compares the nominal link budgets for lander to Earth DTE, lander to Titan orbiter, and Titan orbiter relay to Earth. Nominal values have been chosen for the link budget based on available information for Cassini-Huygens and Dragonfly missions, and the calculations are only accurate to an order of magnitude. It is seen that for lander DTE using X-Band, the nominal data rate is 2 kbps. If the lander instead communicated using X-band to the nearby Titan orbiter, even with one-third of the power, data rates as high as 10 Mbps can be achieved. For the Titan orbiter relay to Earth using Ka-Band, data rates as high as 200 kbps can be achieved. In addition, unlike the lander which looses view of Earth every 8 days, the orbiter will have frequent overhead passes which are visible from the lander. The lander will have the opportunity to communicate to the orbiter for a few minutes during every pass. This implies the lander can uplink a large volume of data during a short pass of only a few minutes, which the orbiter will store on board its memory and then play back to Earth at its lower data rate in between its science operations.

Table 2. Link Budgets for DTE and Relay to Earth via Orbiter

|  | **Lander to Earth DTE** | **Lander to Titan Orbiter Uplink** | **Titan Orbiter relay to Earth** |
|---|---|---|---|
| Frequency, GHz | 8.425 (X-Band) | 7.70 (X-Band) | 32.00 (Ka-Band) |
| Transmitter Power, W | 30 | 10 | 30 |
| Transmitter Power, dB | 14.77 | 10.0 | 14.77 |
| Transmitter Loss, dB | -1.0 | -1.0 | -1.0 |
| S/C Circuit Loss, dB | -0.2 | -0.2 | -0.2 |
| Transmitter Gain, dBi | 30 | 30 | 60 |
| Receiver Gain, dBi | 80.0 | 32.0 | 80.0 |
| System Noise Temp, K | 40.5 | 230 | 40.5 |
| System Noise Temp, dBK | -16.07 | -23.62 | -16.07 |
| Link Distance, km | 1.5E9 | 10,000 | 1.5E9 |
| Free Space Loss, dB | -294.47 | -190.17 | -306.07 |
| Atmospheric Loss, dB | -0.35 | -0.05 | -0.35 |
| Other Losses, dB | -0.54 | -3.3 | -0.54 |
| Boltzmann's const, dB | +228.6 | +228.6 | +228.6 |
| **Data rate, kbps** | **2.0** | **10,000** | **200** |
| Data rate, dBhz | -33.01 | -70.0 | -53.01 |
| **Available Eb/N0** | 6.72 | 11.26 | 5.46 |
| **Required Eb/N0** | 0.31 | 2.55 | 0.31 |
| **Eb/N0 Margin** | 6.41 | 8.71 | 5.15 |

A simple analysis is presented to compare the data volumes from DTE and a relay orbiter. Figure 7 shows a nominal transmit duty cycle, the lander transmits data during 4 of the 8 days the Earth is above the horizon.

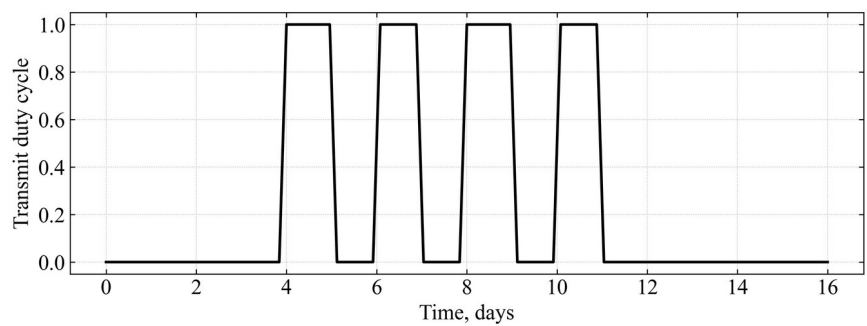

Figure 7. Nominal lander DTE transmit duty cycle.

Figure 8 shows the cumulative data volume for the DTE architecture with the nominal data rate of 2 kbps over the course of a Titan day (16 Earth days). The total data volume is approximately 0.7 Gbits. Figure 9 shows the elevation of the orbiter in the Titan sky as seen from the lander during a Titan day. Once every few days, the orbiter has a series of passes over the lander during in which it transmits data at 10 Mbps to the orbiter which stores it on board. Figure 10 shows the cumulative data volume for the orbiter to Earth relay at 200 kbps, assuming a transmit duty cycle during 40% of the orbit, and when Earth is not in Titan shadow. The total data volume relayed by the orbiter is about 80 Gbits, over a 100 times that possible with DTE from the lander. Over the period of a four year mission (90 Titan Days), the total data volume adds up to 7 Tbits, which is comparable science return to that of a Flagship mission.

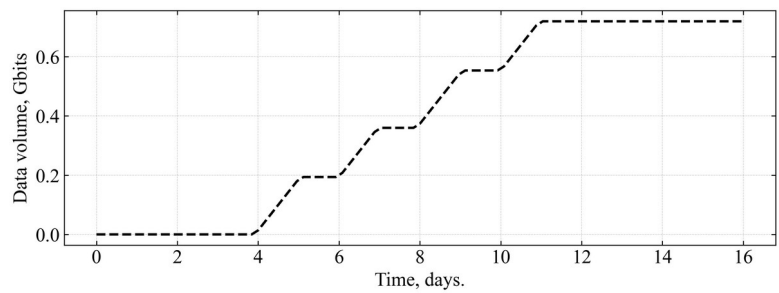

Figure 8. Nominal lander to Earth DTE data volume.

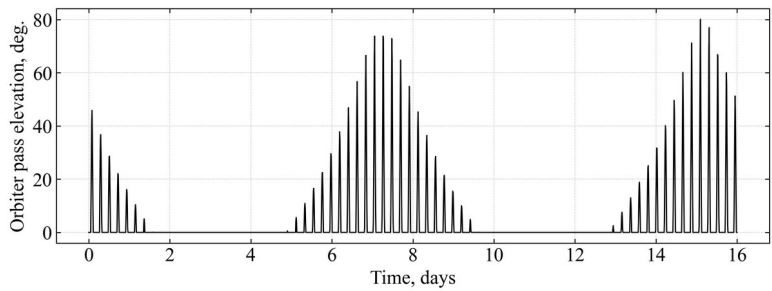

Figure 9. Orbiter pass elevation from landing site.

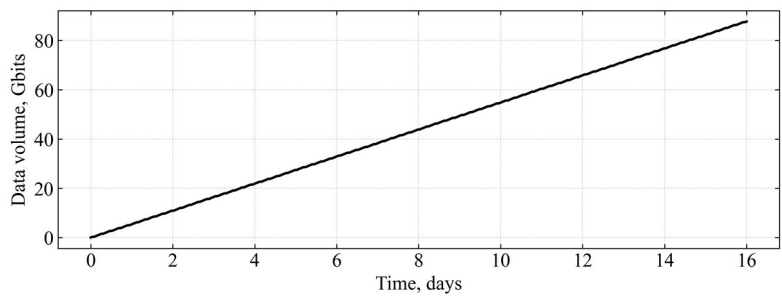

Figure 10. Nominal orbiter to Earth data volume.

## V. APPLICATIONS TO FUTURE MISSIONS

The paper has established two conclusions: 1) The use of ADEPT drag modulation aerocapture enables direct insertion of a large orbiter around Titan without the need for complex control systems used in blunt-body aeroshells. The maneuver provides 5+ km/s of ΔV by using Titan's thick atmosphere, with almost no propellant. 2) Compared to a lander using DTE communication, a Titan orbiter which can relay the data from the lander can increase the data volume (a proxy for the science return) by over a factor of 100. This section explores the implications of these findings for future Titan exploration, given the current science priorities and the funding scenario for future missions.

The most high priority objective for Titan exploration post-Dragonfly will be a Titan orbirer which will perform global mapping of Titan and complement Dragonfly's local and regional in-situ investigation. This is comparable in some ways to how global mapping by the Mars Global Surveyor transformed our understanding of Mars from orbit. The 2023-2032 Planetary Science Decadal Survey has included a Titan Orbiter in its recommendations for the New Frontiers (NF) 6 mission themes. As discussed earlier, getting an orbiter around Titan requires significant ΔV. Without the use of aerocapture, the spacecraft would likely need a SEP stage, use chemical propulsion to capture around Saturn, perform multiple Titan flybys, and then aerobraking at Titan to achieve its orbit. The required ΔV drives the size of the proulsion system and the overall spaceccraft mass, and is the largest driver of mission cost. Aerocapture reduces the ΔV from 5+ km/s to just a few hundered m/s. Aerocapture using blunt-body aeroshells come with complex control systems and cost in the range of $2B making them incompatible with the NF cost cap. Drag modulation aerocapture removes the need for complex flight control systems, and thus have the potential to significantly reduce the cost such that it may be possible to fit a Titan Orbiter within the $1B NF cost cap. Previous studies have shown that it is possible to fit outer planet orbiter missions using propulsive insertion at Uranus in NF, though orbit insertion is easier at Uranus than Titan [35, 36]. Additional studies are required to estimate the cost of an drag modulation aerocapture Titan mission and if it fits within NF, but if it can be realized then the Titan Orbiter will make a strong contender for a future NF mission. NASA aims to conduct two NF missions per decade, though this goal has been challenging to realize. Even though it will be well beyond Dragonfly's planned mission life, it is possible Dragonfly continues to operate when a future NF orbiter arrives at Titan. Since Dragonfly does not have any consumables that limit its mission life or significant mechanical wear, and the degradation of the RTG power can be offset by simply allowing a longer charging time for the battery, Dragonfly may continue to operate for several years beyond its planned mission. If a Titan orbiter arrives before its end of mission, it becomes possible to perform coordinated investigations from orbit and the ground, as well as greatly increase the data returned by the Dragonfly mission.

Flagship class misison concepts continue to be proposed for Titan [34]. The 2023-2032 Decadal Survey has recommended the Uranus Orbiter and Probe (UOP) as the Flagship mission for the next decade, and an Enceladus-Orbilander as the next highest priority. Given this scenario, it appears Titan Flagship missions may not be feasible in the next two decades. However, it is possible to do Flagship-class science at Titan using two New Frontiers class misisons. For example, a Dragonfly-like lander and an ADEPT drag modulation aerocapture orbiter if it each fits within the NF cost cap, can in coordination to provide the science return equivalent to a Flagship-class mission. While the lander undertakes the local in-situ measurements, the orbiter will perform global measurements thus complementing each other's scientific investigations. In addition, the orbiter acting as a relay will allow the lander to return far more science data than it could using DTE. From the perspective of science return, the ability of the orbiter to multiply the lander data volume by a factor of 100 makes it a force multiplier for bits returned and hence science return per dollar. This could potentially allow millions of high resolution pictures from Titan to be returned over the course of a mission, making Titan's surface as familiar to the public as Mars. Since NF missions are more frequent than Flagships, there exists more opportunities to undertake such an approach within the next two decades. There also exists the possibility of international collaboration, as NASA and ESA for example can each fly one of the lander or orbiter, and together they can perform synergetic science and overcome the limitations of DTE for the lander.

## VI. CONCLUSIONS

Dragonfly, a single element mission which fits within the New Frontiers cost cap will arrive at Titan in 2034 and land in the equatorial sand dunes to perform in-situ investigations of the organic materials on the surface. However, its detailed investigations will be limited to region within its short flight range. The big gaps in our understanding of Titan's global topography, climate, and upper atmospheric chemistry which can only investigated from an orbiter around Titan will remain to be addressed by a future orbiter mission. Past studies have shown that it is challenging to fit a Titan orbiter within the NF cost cap. ADEPT drag modulation aerocapture, a simple flight control technique which uses Titan's thick atmosphere to provide the large $\Delta V$ required for orbit insertion may enable a Titan orbiter to fit within the NF cost cap. While a Flagship mission consisting of an orbiter and a lander operating together is ideal for both synergy of science and return of data from the lander, current funding priorities preclude such a mission in the next two decades. The present study explored options to do Flagship-class science at Titan using two New Frontiers class misisons. A Dragonfly-like lander and an ADEPT drag modulation aerocapture orbiter if it each fits within the NF cost cap, can in coordination to provide the science return equivalent to a Flagship-class mission.

# DATA AVAILABILITY

All the results were created using the Aerocapture Mission Analysis Tool (AMAT v2.2.22). Jupyter Notebooks to reproduce the results are available at https://github.com/athulpg007/AMAT/tree/master/examples/titan-relay-orbiter

# REFERENCES

[1] Dyer JW, "Pioneer Saturn," *Science,* Vol. 207, No. 4429, 1980, pp. 400-401. https://doi.org/10.1126/science.207.4429.400

[2] Stone EC, Miner ED, "Voyager 1 encounter with the Saturnian system," *Science*, Vol. 212, No. 4491, 1981, pp. 159-163. https://doi.org/10.1126/science.212.4491.159

[3] Lebreton JP, et al., "An overview of the descent and landing of the Huygens probe on Titan," *Nature,* Vol. 438, No. 7069, 2005, pp. 758-764. https://doi.org/10.1038/nature04347

[4] Stofan ER, et al. "The lakes of Titan," *Nature,* Vol. 445, No. 7123, 2007, pp. 61-64. https://doi.org/10.1038/nature05438

[5] Lorenz RD, et al., "The sand seas of Titan: Cassini RADAR observations of longitudinal dunes," *Science,* Vol. 312, No. 5774, 2006, pp. 724-727. https://doi.org/10.1126/science.1123257

[6] Hayes AG et al., "A post-Cassini view of Titan's methane-based hydrologic cycle," *Nature Geoscience,* Vol. 11, No. 5, 2018, pp. 306-313. https://doi.org/10.1038/s41561-018-0103-y

[7] Mackenzie SM, et al., "Titan: Earth-like on the outside, ocean world on the inside," *The Planetary Science Journal,* Vol. 2, No. 3, 2021, pp. 112. https://doi.org/10.3847/PSJ/abf7c9

[8] He C, Smith MA, "Identification of nitrogenous organic species in Titan aerosols analogs: Implication for prebiotic chemistry on Titan and early Earth," *Icarus*, Vol. 238, 2014, pp. 86-92. https://doi.org/10.1016/j.icarus.2014.05.012

[9] Elliott JO, Reh K, Spilker T, "Concept for Titan exploration using a radioisotopically heated montgolfiere," *2007 IEEE Aerospace Conference,* IEEE, 2007, pp 1-11. https://doi.org/10.1109/AERO.2007.352717

[10] Lorenz, RD et al, "Titan explorer: A NASA flagship mission concept," *AIP Conference Proceedings,* Vol. 969. No. 1, American Institute of Physics, 2008, pp. 380-387. https://doi.org/10.1063/1.2844991

[11] Reh K, "Titan Saturn System Mission," *2009 IEEE Aerospace Conference,* IEEE, 2009, pp. 1-8. https://doi.org/10.1109/AERO.2009.4839316

[12] Stofan E, "Time-the titan mare explorer," *2013 IEEE Aerospace Conference,* IEEE, 2013, pp 1-10. https://doi.org/10.1109/AERO.2013.6497165

[13] Barnes JW et al., "AVIATR—Aerial Vehicle for In-situ and Airborne Titan Reconnaissance: A Titan airplane mission concept," *Experimental Astronomy,* Vol. 33, 2012, pp. 55-127. https://doi.org/10.1007/s10686-011-9275-9

[14] Barnes, JW et al. "Science goals and objectives for the Dragonfly Titan rotorcraft relocatable lander," *The Planetary Science Journal,* Vol. 2, No. 4, 2021, pp. 130. https://doi.org/10.3847/PSJ/abfdcf

[15] Lorenz, RD et al., "Selection and characteristics of the Dragonfly landing site near Selk crater, Titan," *The Planetary Science Journal* Vol. 2, No.1, 2021, pp. 24.
https://doi.org/10.3847/PSJ/abd08f

[16] Barnes JW et al., "New Frontiers Titan Orbiter," *Bulletin of the American Astronomical Society,* Vol. 53. No. 4, 2021, pp. 317.
https://doi.org/10.3847/25c2cfeb.4c2df948

[17] Lockwood MK, "Titan aerocapture systems analysis," *39th AIAA/ASME/SAE/ASEE Joint Propulsion Conference and Exhibit,* 2003, pp. 4799.
https://doi.org/10.2514/1.A32589

[18] Putnam ZR and Braun RD, "Drag-modulation flight-control system options for planetary aerocapture," *Journal of Spacecraft and Rockets,* Vol. 51, No. 1, 2014, pp. 139-150.
https://doi.org/10.2514/1.A32589

[19] Austin A et al., "Enabling and Enhancing Science Exploration Across the Solar System: Aerocapture Technology for SmallSat to Flagship Missions," *Bulletin of the American Astronomical Society,* Vol. 53, No. 4, 2021, pp. 057.
https://doi.org/10.3847/25c2cfeb.4b23741d

[20] Girija AP, Lu Y, and Saikia SJ, "Feasibility and mass-benefit analysis of aerocapture for missions to Venus," *Journal of Spacecraft and Rockets,* Vol. 57, No. 1, 2020, pp. 58-73.
https://doi.org/10.2514/1.A34529

[21] Wercinski, P, "Adaptable Deployable Entry and Placement Technology (ADEPT) Enabling Future Science Missions," *NASA Tech Showcase,* 2023.
https://ntrs.nasa.gov/citations/20220019227

[22] Dutta S et al., "Adaptable Deployable Entry and Placement Technology Sounding Rocket One Modeling and Reconstruction," *Journal of Spacecraft and Rockets,* Vol. 59, No. 1, 2022, pp. 236-259.
https://doi.org/10.2514/1.A35090

[23] Girija AP, Saikia SJ, and Longuski JM, "Aerocapture: Enabling Small Spacecraft Direct Access to Low-Circular Orbits for Planetary Constellations," *Aerospace,* Vol. 10, No. 3, 2023, pp. 271.
https://doi.org/10.3390/aerospace10030271

[24] Dutta S et al., "Aerocapture as an Enhancing Option for Ice Giants Missions," *Bulletin of the American Astronomical Society,* Vol. 53, No. 4, 2021, pp. 046.
https://doi.org/10.3847/25c2cfeb.e8e49d0e

[25] Girija AP et al., "Feasibility and performance analysis of neptune aerocapture using heritage blunt-body aeroshells," *Journal of Spacecraft and Rockets* Vol. 57, No. 6, 2020, pp. 1186-1203.
https://doi.org/10.2514/1.A34719

[26] Girija AP, "A Flagship-class Uranus Orbiter and Probe mission concept using aerocapture," *Acta Astronautica* Vol. 202, 2023, pp. 104-118.
https://doi.org/10.1016/j.actaastro.2022.10.005

[27] Strauss WD et al. "Aerocapture Trajectories for Earth Orbit Technology Demonstration and Orbiter Science Missions at Venus, Earth, Mars, and Neptune," *AAS/AIAA Astrodynamics Specialist Conference,* 2021, pp 1-22.
https://hdl.handle.net/2014/54292

[28] Scott CJ et al., "Preliminary interplanetary mission design and navigation for the Dragonfly New Frontiers mission concept," *AAS/AIAA Astrodynamics Specialist Conference,* 2018, pp 1-21.
https://hdl.handle.net/2014/48626

[29] Strange N et al., "Mission design for the titan saturn system mission concept," *AAS/AIAA Astrodynamics*

*Specialist Conference*, 2009, pp. 1-16.

[30] Girija AP, "A Systems Framework and Analysis Tool for Rapid Conceptual Design of Aerocapture Missions," Ph.D. Dissertation, Purdue University Graduate School, 2021.
https://doi.org/10.25394/PGS.14903349.v1

[31] Girija AP et al., "Aerocapture Performance Analysis for a Neptune Mission Using a Heritage Blunt-Body Aeroshell," *AAS/AIAA Astrodynamics Specialist Conference,* 2019, pp 1-21.
https://doi.org/10.31224/osf.io/bf3du

[32] Girija AP et al. "AMAT: A Python package for rapid conceptual design of aerocapture and atmospheric Entry, Descent, and Landing (EDL) missions in a Jupyter environment," *Journal of Open Source Software,* Vol. 6, No. 67, 2021, pp. 3710.
https://doi.org/10.21105/joss.03710

[33] Girija AP et al. "Quantitative assessment of aerocapture and applications to future solar system exploration." *Journal of Spacecraft and Rockets,* Vol. 59, No. 4, 2022, pp. 1074-1095.
https://doi.org/10.2514/1.A35214

[34] Nixon C et al., "The Science Case for a Titan Flagship-class Orbiter with Probes," *Bulletin of the American Astronomical Society,* Vol. 53, No. 4, 2021, pp. 325.
https://doi.org/10.3847/25c2cfeb.bc2b9583

[35] Jarmak S et al., "QUEST: A New Frontiers Uranus orbiter mission concept study," *Acta Astronautica,* Vol. 170, 2020, pp. 6-26.
https://doi.org/10.1016/j.actaastro.2020.01.030

[36] Cohen I et al., "New Frontiers-class Uranus Orbiter: Exploring the feasibility of achieving multidisciplinary science with a mid-scale mission," *Bulletin of the American Astronomical Society,* Vol. 53, No. 4, 2021, pp. 323.
https://doi.org/10.3847/25c2cfeb.262fe20d